\documentclass[prb,superscriptaddress,twocolumn,nofootinbib]{revtex4}

\usepackage{graphics} 
\usepackage{epsfig} 

\begin{document}   
\title{Comparative study of the growth of sputtered aluminum oxide films on organic and inorganic substrates}

\author{Stefan Sellner}
\affiliation{Max-Planck-Institut f\"ur Metallforschung, Heisenbergstr. 3,
70569 Stuttgart, Germany}
\affiliation{Institut f\"ur Theoretische und Angewandte Physik, Universit\"at 
Stuttgart, Pfaffenwaldring 57, 70550 Stuttgart, Germany}
\affiliation{Institut f\"ur Angewandte Physik, Universit\"at T\"ubingen,
  Auf der Morgenstelle 10, 72076 T\"ubingen, Germany} 

\author{Alexander Gerlach}
\affiliation{Institut f\"ur Angewandte Physik, Universit\"at T\"ubingen,
  Auf der Morgenstelle 10, 72076 T\"ubingen, Germany} 
\affiliation{Physical and Theoretical Chemistry Laboratory, Oxford University,
  South Parks Road, Oxford OX1 3QZ, United Kingdom}

\author{Stefan Kowarik} 
\affiliation{Institut f\"ur Angewandte Physik, Universit\"at T\"ubingen,
  Auf der Morgenstelle 10, 72076 T\"ubingen, Germany} 
\affiliation{Physical and Theoretical Chemistry Laboratory, Oxford University,
  South Parks Road, Oxford OX1 3QZ, United Kingdom}

\author{Frank Schreiber}
\email[corresponding author:]{frank.schreiber@uni-tuebingen.de}
\affiliation{Institut f\"ur Angewandte Physik, Universit\"at T\"ubingen,
  Auf der Morgenstelle 10, 72076 T\"ubingen, Germany} 
\affiliation{Physical and Theoretical Chemistry Laboratory, Oxford University,
  South Parks Road, Oxford OX1 3QZ, United Kingdom}
  
\author{Helmut Dosch}
\affiliation{Max-Planck-Institut f\"ur Metallforschung, Heisenbergstr. 3,
  70569 Stuttgart, Germany}
\affiliation{Institut f\"ur Theoretische und Angewandte Physik, Universit\"at
  Stuttgart, Pfaffenwaldring 57, 70550 Stuttgart, Germany}

\author{Stephan Meyer}
\author{Jens Pflaum}
\affiliation{III. Physikalisches Institut, Universit\"at Stuttgart,
  Pfaffenwaldring 57, 70550 Stuttgart, Germany}

\author{Gerhard Ulbricht}
\affiliation{Max-Planck-Institut f\"ur Festk\"orperforschung, 
Heisenbergstr.1, 70569 Stuttgart, Germany}

\begin{abstract}
  We present a comparative study of the growth of the technologically highly
  relevant gate dielectric and encapsulation material aluminum oxide in
  inorganic and also organic heterostructures. Atomic force microscopy studies
  indicate strong similarities in the surface morphology of aluminum oxide
  films grown on these chemically different substrates. In addition, from
  X-ray reflectivity measurements we extract the roughness exponent $\beta$ of
  aluminum oxide growth on both substrates. By renormalising the aluminum
  oxide roughness by the roughness of the underlying organic film we find good
  agreement with $\beta$ as obtained from the aluminum oxide on silicon oxide
  ($\beta = 0.38 \pm 0.02$), suggesting a remarkable similarity of the
  aluminum oxide growth on the two substrates under the conditions
  employed.\\[.1em]
\end{abstract}

\date{\today}
\maketitle

\section{Introduction}

Functional thin films receive growing attention in different fields such as
microelectronics, optics and coating technology. Because of its extraordinary
mechanical, electrical, thermal and optical properties aluminum oxide has
become an important thin film material for various applications.  The large
band gap of aluminum oxide, for example, facilitates its use in magnetic
tunnel junctions\cite{rippard_prl02}, the low thermal conductivity on the
other hand makes it a very suitable material for thermal barrier coatings as
they are used, \textit{e.g.}, in gas-turbine engines.\cite{padture_science02}
Ultrathin and well-ordered aluminum oxide layers on metal
substrates\cite{stierle_science04} exhibit catalytic activity, whereas
amorphous films yield highly stable dielectric encapsulation layers -- an
application especially important in the emerging field of organic
semiconductors, where device encapsulation is necessary to guarantee a long
term stability.  

For this purpose different approaches using transparent aluminum oxide films
were shown to fulfill the technological
requirements\cite{park_etrij05,chwang_apl03} and thus turn the vision of
flexible displays to a more realistic prospect.  Yet, the growth processes of
such films represent a fundamental challenge with direct impact on device
performance, \textit{e.g.}, on the breakthrough voltage in organic
field-effect transistors.\cite{voigt_mseb04} One critical parameter is the
evolution of the film roughness with increasing film thickness.  In the theory
of growth processes scaling theories for the surface morphology and dynamics
of a growing film have become a very successful
concept.\cite{barabasi-stanley_bk,krug_advp97,pimpinelli_bk} In the so-called
dynamic scaling regime and for a constant deposition rate the root mean square
(rms) surface roughness $\sigma$ of a film scales with the film thickness
$L$,\cite{family_jpa85,family_pa90}
\begin{equation}
  \sigma \propto L^\beta,
  \label{eq:scalinglaw}
\end{equation}
where the growth exponent $\beta$ depends on the mechanism of the film growth.
The dynamic scaling formalism has been applied to different theoretical models
of growing
interfaces\cite{family_jpa85,family_pa90,family_jpa86,kardar_prl86,jullien_jpa85,wolf_epl90,krug_prl91,lai_prl91,tang_prl91}
and experimental studies show that depending on the deposition method and on
the materials one typically obtains $0.2 \leq \beta \leq
1$.\cite{he_prl92,ernst_prl94,thompson_prb94,you_prl93,jun_epl98,collins_prl94,duerr_prl03}

In this paper we present a study on the growth of sputtered aluminum oxide
films deposited on two very different surfaces, namely silicon oxide and films
of the organic semiconductor diindenoperylene (DIP). The structure of DIP
films has been studied in detail\cite{duerr_apl02,kowarik_prl06} and DIP has
already served as organic model system for studies on metal deposition
\cite{duerr_am02} and encapsulation
methods.\cite{sellner_am04,sellner_LPaper,meyer_submitted} Using two
complementary techniques, \textit{i.e.\/} atomic force microscopy (AFM) and
X-ray diffraction, both the surface morphology and the roughness evolution was
studied. The roughness exponent $\beta$ for sputtered aluminum oxide films
deposited on silicon oxide and organic substrates could be determined.

\section{Experimental Details}
\label{sec:exp}

Silicon wafers [Si(100)] with a native oxide layer were used as a substrate.
Before deposition of the organic films the substrates were cleaned in an
ultrasonic bath with acetone and ethanol and outgased in the ultra-high vacuum
(UHV) chamber at 700\,$^\circ$C for 12\,hours. The DIP films were prepared by
organic molecular-beam deposition under UHV conditions as described
elsewhere.\cite{duerr_apl02}

The aluminum oxide films were prepared by radio frequency magnetron sputtering
in a dedicated high-vacuum chamber (base pressure $3 \times
10^{-5}$\,Pa).\cite{sellner_LPaper} To avoid oxidation of the organic film
pure argon was used as sputter gas. The sputtering unit (AJA International
ST30) was equipped with an aluminum oxide target and operated under an argon
atmosphere of $0.2 - 0.3$\,Pa. Regarding the oxygen content this leads to an
understoichiometric target after some sputtering cycles which had been
overcome by regenerating the target after each deposition in an argon/oxygen
atmosphere ($p(\mathrm{Ar})= 0.5\,$Pa / $p(\mathrm{O_2})= 0.2\,$Pa). The
gases used had a purity of 99.999\%. Despite the low sputtering power of $120
- 200$\,W the substrates were water cooled during the deposition
($T_\mathit{substr.}= -10\,^\circ$C).  The deposition rate of
$\sim$\,7\,\AA/min was determined by a quartz crystal microbalance which was
calibrated beforehand by X-ray reflectivity measurements on as-prepared films.
The stoichiometry of sputtered aluminum oxide films was determined by
Rutherford backscattering spectroscopy (RBS).\footnote{The measurements were
  carried out with He$^+$ ions of $1$\,MeV at the Dynamitron in Stuttgart. The
  RBS-chamber has an IBM-geometry (\textit{i.e.\/}, the detector is located at
  $\theta$=165$^\circ$ scattering angle in the same plane as the beam and the
  normal to the sample) with a detector resolution of 14\,keV FWHM.} The
samples studied here had a typical Al/O ratio of 0.63, \textit{i.e.\/} close
to the stoichiometry of Al$_2$O$_3$.  The argon content was below 1\,at.\% for
all samples. We note that after the aluminum oxide sputtering process no
significant decomposition of the crystalline structure of the DIP substrate
(except for the topmost one or two monolayers) was
observed.\cite{sellner_LPaper}

After preparation of the oxide films, the samples were analyzed by means of
atomic force microscopy (AFM) and specular X-ray diffraction. The AFM
measurements were performed in contact mode under UHV conditions. The X-ray
diffraction measurements were made with a laboratory source (with
Cu$_{K_\alpha}$: $\lambda = 1.54$\,\AA) and at the ANKA synchrotron radiation
source in Karlsruhe (with $\lambda = 1.08$\,\AA).  

\section{Results}
\subsubsection*{Surface morphology of Al$_2$O$_3$/SiO$_x$ and Al$_2$O$_3$/DIP
  -- AFM}

\begin{figure}[htbp]
  \centering
    \includegraphics[width=8.5cm]{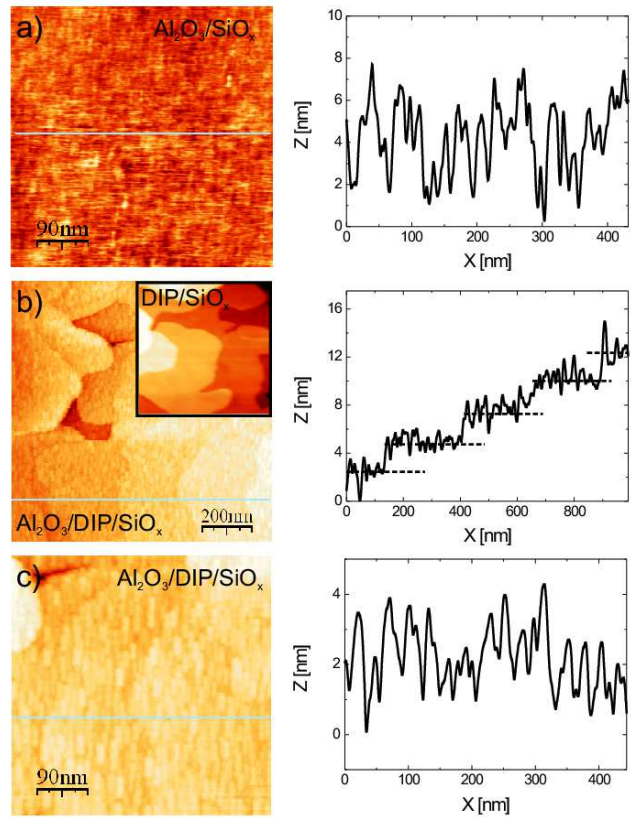}
    \caption{Topographical AFM images (contact mode) with line scans of a 
      174\,\AA\/ thick aluminum oxide film deposited on silicon oxide (a) and
      of a 681\,\AA\/ thick aluminum oxide film on DIP (b). The inset in (b)
      shows the typical topography of the organic film before aluminum oxide
      deposition. (c) Close-up AFM image from (b) showing the morphology of
      the aluminum oxide film on a single DIP terrace.}
  \label{fig:CompAFM}
\end{figure}

After the sputtering process the surface morphology of the aluminum oxide
films was investigated by contact mode AFM. Fig.~\ref{fig:CompAFM}(a) shows a
typical image of a $\sim$174\,\AA\/ thick sputtered aluminum oxide film on
silicon oxide with a line scan of the sample topography. The relatively smooth
film surface exhibits a grainy morphology with a mean distance of its grains
of about 14.5\,nm. For thicker aluminum oxide films on silicon oxide a similar
morphology was found.
Fig.~\ref{fig:CompAFM}(b) shows an AFM image of a 681\,\AA\/ thick aluminum
oxide film deposited under similar sputtering conditions on top of a DIP film
of 317\,\AA\/ thickness. The inset shows a contact mode AFM image of an
uncapped DIP film with its characteristic topography with terraces of
monomolecular (ca.\ 16.5\,\AA) step height. The corresponding line scan reveals
the surface morphology of aluminum oxide/DIP.

For the Al$_2$O$_3$/DIP system (Fig.~\ref{fig:CompAFM}(b)) the terraced
structure of the underlying DIP film can still be recognized which implies
that the Al$_2$O$_3$ surface roughness exhibits a certain degree of
correlation with the DIP surface roughness. A closer look at the morphology of
the aluminum oxide layer on a DIP terrace (Fig.~\ref{fig:CompAFM}(c)) exhibits
a granular structure as could already be seen on the Al$_2$O$_3$/SiO$_x$
system (Fig.~\ref{fig:CompAFM}(a)).  The aluminum oxide film thus reflects
some features of the underlying substrates -- the relatively flat native
silicon oxide and the terraced DIP -- in addition to its 'inherent'
graininess.

\subsubsection*{Roughness evolution of Al$_2$O$_3$/SiO$_x$ and Al$_2$O$_3$/DIP
  -- X-ray reflectivity}

\begin{figure}[htbp]
  \centering
    \includegraphics[width=8.5cm]{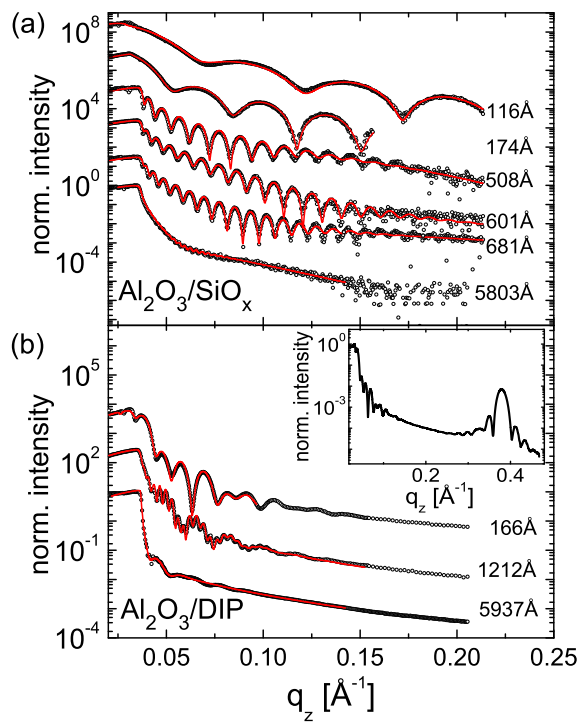}
    \caption{X-ray reflectivities of sputtered aluminum oxide layers of
      different thickness (a) on silicon oxide substrates and (b) on DIP
      films. The data in (a) were measured at a laboratory source (Cu
      K$_{\alpha}$) and the data in (b) were taken at the ANKA synchrotron
      facility in Karlsruhe at $E=11.5$\,keV. The inset in (b) shows the 
      reflectivity of the 166\,\AA\/ thick Al$_2$O$_3$/DIP film including 
      the first order DIP Bragg reflection at $q_z=0.38$\,\AA $^{-1}$.}
  \label{fig:Aloxlayers}
\end{figure}

From the X-ray reflectivity measurements the out-of-plane structure is probed
and information on the film thickness, the electron density and interface
roughness can be extracted. Aluminum oxide films of thicknesses ranging from
ca.\ 116\,\AA\/ to 5800\,\AA\/ were prepared on both substrates. The
experimental data of X-ray reflectivity measurements and fits using the
Parratt formalism \cite{parratt_pr54} are displayed in
Fig.~\ref{fig:Aloxlayers}(a) for Al$_2$O$_3$/SiO$_x$ and in
Fig.~\ref{fig:Aloxlayers}(b) for Al$_2$O$_3$/DIP.  The X-ray reflectivity
curves are offset for clarity. The specular signal was obtained by subtracting
the off-specular diffuse signal from the measured intensity. The inset in
Fig.~\ref{fig:Aloxlayers}(b) shows the reflectivity of the 166\,\AA\/ thick
Al$_2$O$_3$/DIP film including the first order DIP Bragg reflection at
$q_z=0.38$\,\AA $^{-1}$.

The specular X-ray reflectivity curves show pronounced thickness oscillations
(Kiessig fringes) indicating well-defined interfaces (in terms of interface
roughness).\footnote{The reflectivity curve for 5800\,\AA\/ aluminum oxide on
  SiO$_x$ in Fig.~\ref{fig:Aloxlayers}(a) shows some oscillations which do not
  correspond to the thickness oscillations of a 5800\,\AA\/ thick aluminum
  oxide film. During preparation of this specific film the sputtering process
  was interrupted to cool down the sputtering target. After restarting
  sputtering the preparation conditions might have changed slightly. The
  reflectivity curve could be fitted by using a model with two aluminum oxide
  films of slightly different electron density but the total film thickness
  being ca.\ 5800\,\AA.} For all films no signature of crystalline aluminum
oxide could be found at higher scattering angles, \textit{i.e.}, at the
position of Bragg reflections of $\alpha$-Al$_2$O$_3$ at $q_z =
2.46$\,\AA$^{-1}$ and $q_z=3.01$\,\AA$^{-1}$. From the fitting of the X-ray
data the film thickness, the electron density and the film-substrate roughness
as well as the film surface roughness was determined.\cite{parratt_pr54}

\section{Analysis and Discussion}

For many systems prepared by different deposition methods the imperfection of
a given layer is transferred fully or partly to the subsequent
layers.\cite{schlomka_prb95} A simple way of taking into account the effect of
vertical correlations between the interfaces in our samples is to
'renormalize' the roughness of the aluminum oxide film by the rms roughness of
the underlying substrate, according to\cite{holy_bk}
\begin{equation}
  \sigma^\mathit{renorm.}_{Al_2O_3} = \sqrt{\sigma_{Al_2O_3}^2 -
  \sigma^2_\mathit{substr.}}\  ,
\label{eq:renormalization}
\end{equation}
where $\sigma_\mathit{substr.}$ corresponds to the roughness of the underlying
film/substrate, \textit{i.e.\/} DIP or SiO$_x$.

\subsubsection*{Aluminum oxide on SiO$_x$}

\begin{figure}[htbp]
  \centering
    \includegraphics[width=8.5cm]{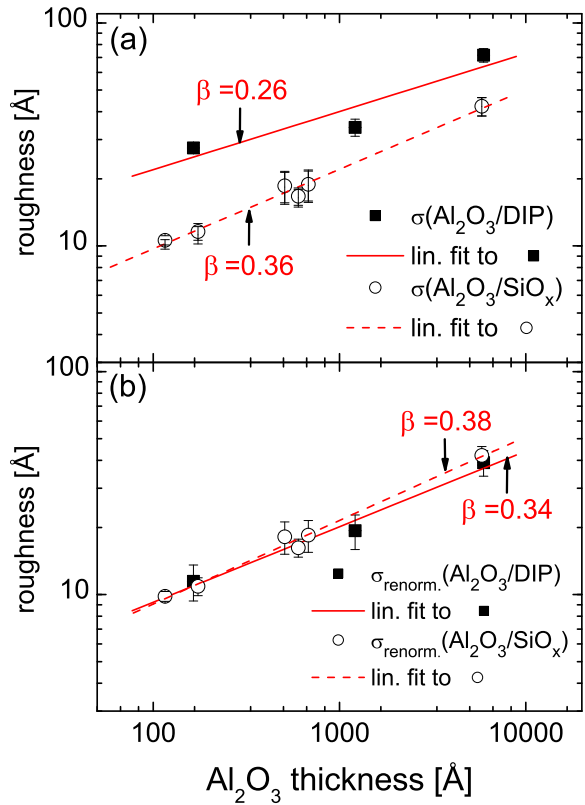}
    \caption[Corrected roughness for Al$_2$O$_3$/DIP compared to the roughness
    of Al$_2$O$_3$/SiO$_x$.]{(a) Roughness $\sigma_{Al_2O_3}$ for
      Al$_2$O$_3$/DIP (filled squares) and for Al$_2$O$_3$/SiO$_x$ (open
      circles) without correcting for the roughness of the underlying
      substrate. (b) Renormalized roughness
      $\sigma^\mathit{renorm.}_{Al_2O_3}$(DIP) for Al$_2$O$_3$/DIP (filled
      squares) compared to the roughness
      $\sigma^\mathit{renorm.}_{Al_2O_3}$(SiO$_x$) of the Al$_2$O$_3$/SiO$_x$
      system (open circles). The scaling behavior of aluminum oxide layers
      deposited on DIP and on SiO$_x$ are in good agreement.}
  \label{fig:Aloxdiproughness}
\end{figure}

In Fig.~\ref{fig:Aloxdiproughness}(a) the roughness evolution of sputtered
aluminum oxide films deposited on silicon oxide is displayed in a $log$-$log$
plot as a function of the film thickness (open circles). The slope of a linear
fit to the data corresponds to a roughness exponent of $\beta = 0.36 \pm
0.05$. The roughness of the silicon oxide substrate was 4\,\AA\/ as determined
from measurements on the clean substrate. When the aluminum oxide roughness is
corrected for the relatively small roughness of the native silicon oxide
substrate using Eq.~(\ref{eq:renormalization}) a roughness exponent of $\beta
= 0.38 \pm 0.02$ is obtained (see Fig.~\ref{fig:Aloxdiproughness}(b)).

This result has to be compared to the wide range of experimental studies of
different materials deposited by different techniques. For instance, for
thermal evaporation of Fe on Fe(001)\cite{he_prl92} $\beta=0.22$ and for vapor
deposited Ag on silicon substrates\cite{thompson_prb94} a scaling exponent of
$\beta=0.26$ has been obtained, while for sputter-deposited Au films on
Si(111) $\beta=0.40$ (at 300\,K) and $\beta=0.42$ (at 200\,K) \cite{you_prl93}
and for sputtered Mo films on Si(111) $\beta=0.42$ was
reported\cite{jun_epl98}.  For aluminum nitride (AlN) films deposited by
reactive-sputtering on Si(100) substrates $\beta=0.37$ was
found\cite{auger_jap05} while for sputtered SnO$_2$ films on glass substrates
a growth exponent of approximately 0.3 is reported.  \cite{lindstroem_tsf01}

The theoretical predictions for $\beta$ depend on the assumptions of the
specific model.\cite{krug_advp97} Obviously, the various $\beta$ values show
that for different materials, substrate temperatures and deposition techniques 
different growth mechanisms are dominating.

\subsubsection*{Aluminum oxide on DIP/SiO$_x$}

The roughness of aluminum oxide films of different thickness deposited on top
of DIP films is plotted in Fig.~\ref{fig:Aloxdiproughness}(a) (filled squares)
and an uncorrected growth exponent of $\beta=0.26 \pm 0.10$ is extracted.
Given the large roughness of the substrate (the DIP film) the renormalization
is essential.

Table~\ref{tab:fitresults} summarizes the thickness and roughness of the
aluminum oxide and DIP films and the renormalized aluminum oxide roughness
$\sigma^\mathit{renorm.}_{Al_2O_3}$. As shown in
Fig.~\ref{fig:Aloxdiproughness}(b) the renormalized aluminum oxide roughness
yields a scaling exponent of $\beta = 0.34 \pm 0.05$\footnote{We note that
  this $\beta$ values arise under the present specific sputtering conditions
  employed in this study, but is not necessarily universal. We rather expect
  it to change with sputtering power, geometry, argon gas pressure and other
  experimental parameters.  } -- a value which is remarkably similar to the
scaling exponent determined for the Al$_2$O$_3$/SiO$_x$ system.  With the
similar morphology of aluminum oxide on silicon oxide substrates
(Fig.~\ref{fig:CompAFM}(a)) and on a single terrace of DIP
(Fig.~\ref{fig:CompAFM}(c)), it appears that the growth and structure of the
aluminum oxide films is similar on both kinds of substrates.

\begin{table}
\begin{tabular}{c|c|c|c||c}
  $L_{Al_2O_3}$~[\AA] & ~$\sigma_{Al_2O_3}$~[\AA] & $L_{DIP}$~[\AA] &
  ~$\sigma_{DIP}$~[\AA] & ~$\sigma_{Al_2O_3}^\mathit{renorm.}$~[\AA] \\
  \hline
  166.4 & 27.5 & 344.6 & 25.0 & 11.5 \\
  1212.4 & 34.0 & 351.5 & 28.0 & 19.3 \\
  5937.6 & 71.7 & 402.0 & 60.2 & 38.9 \\
\end{tabular}
\caption{Results obtained by analyzing the X-ray reflectivity data on
  Al$_2$O$_3$/DIP using the Parratt formalism. The  aluminum oxide roughness is
  corrected by the roughness of the underlying DIP film (see
  Eq.~\ref{eq:renormalization}).}
\label{tab:fitresults}
\end{table}

We note that at least in the initial stage $\sigma_{DIP}\gg
\sigma^\mathit{renorm.}_{Al_2O_3}$, \textit{i.e.\/} that the 'starting
roughness' provided by DIP is the dominating contribution to the aluminum
oxide roughness. Because of the pronounced and well-developed terrace
structure of DIP the renormalization yields the small \textit{'local'}
roughness ($\sigma^\mathit{renorm.}_{Al_2O_3}$) on top of a given terrace,
whereas the \textit{'global'} roughness ($\sigma_{Al_2O_3}$) of aluminum oxide
contains the 'terrace-to-terrace' contribution of the underlying DIP film.
Thus, at least for not too thick films, the renormalization procedure appears
to be a sensible approach.  The remarkable observation is the similarity in
the roughness exponent $\beta$ despite the very different chemical nature of
the two substrates and the substantially lower surface energy of DIP compared
to silicon oxide.

\section{Summary}

We have studied the structure and morphology of aluminum oxide films deposited
on silicon oxide and organic films of DIP. From the analysis of the X-ray
reflectivity measurements we found a roughness exponent of $\beta=0.38 \pm
0.02$ for aluminum oxide films on silicon oxide.  The growth exponent $\beta$
was also determined for sputter deposited aluminum oxide films on DIP films.
The simple renormalization approach of Eq.~(\ref{eq:renormalization}) works
remarkably well. After renormalizing $\sigma_{Al_2O_3}$ in the Al$_2$O$_3$/DIP
system a similar $\beta$-exponent of 0.34 as for the Al$_2$O$_3$/SiO$_x$
system ($\beta =0.38$) was obtained.  The similar growth exponents $\beta$ and
the AFM images of the Al$_2$O$_3$/SiO$_x$ and Al$_2$O$_3$/DIP systems suggest
that the growth and local structure of aluminum oxide exhibit similarities
despite the different chemical nature of the substrates.

\paragraph*{\textbf{Acknowledgements}}
We acknowledge support by the Deutsche Forschungsgemeinschaft (DFG) within the
Focus Programme on organic field effect transistors and support from the
Engineering and Physical Sciences Research Council \mbox{(EPSRC)}. We are
grateful to the FZ Karlsruhe and the ANKA synchrotron, to N.\ Kasper for his
technical assistance and to H.\ Paulus and W.\ Bolse from the Institut f\"ur
Strahlenphysik for the RBS measurements.




\begin{thebibliography}{10}
\expandafter\ifx\csname bibnamefont\endcsname\relax
  \def\bibnamefont#1{#1}\fi
\expandafter\ifx\csname bibfnamefont\endcsname\relax
  \def\bibfnamefont#1{#1}\fi
\expandafter\ifx\csname url\endcsname\relax
  \def\url#1{\texttt{#1}}\fi
\expandafter\ifx\csname urlprefix\endcsname\relax\def\urlprefix{URL }\fi
\providecommand{\bibinfo}[2]{#2}
\providecommand{\eprint}[2][]{\url{#2}}


\bibitem{rippard_prl02}
  \bibinfo{author}{\bibfnamefont{W.~H.}~\bibnamefont{Rippard}},
  \bibinfo{author}{\bibfnamefont{A.~C.}~\bibnamefont{Perrella}},
  \bibinfo{author}{\bibfnamefont{F.~J.}~\bibnamefont{Albert}},
  \bibnamefont{and}
  \bibinfo{author}{\bibfnamefont{R.~A.}~\bibnamefont{Buhrman}},
  \bibinfo{journal}{Phys. Rev. Lett.} \textbf{\bibinfo{volume}{88}}
  (\bibinfo{year}{2002}) \bibinfo{pages}{046805}.
   
 \bibitem{padture_science02}
   \bibinfo{author}{\bibfnamefont{N.~P.}~\bibnamefont{Padture}},
   \bibinfo{author}{\bibfnamefont{M.}~\bibnamefont{Gell}}, \bibnamefont{and}
   \bibinfo{author}{\bibfnamefont{E.~H.}~\bibnamefont{Jordan}},
   \bibinfo{journal}{Science} \textbf{\bibinfo{volume}{296}}
   (\bibinfo{year}{2002}) \bibinfo{pages}{280} .
   
 \bibitem{stierle_science04}
   \bibinfo{author}{\bibfnamefont{A.}~\bibnamefont{Stierle}},
   \bibinfo{author}{\bibfnamefont{F.}~\bibnamefont{Renner}},
   \bibinfo{author}{\bibfnamefont{R.}~\bibnamefont{Streitel}},
   \bibinfo{author}{\bibfnamefont{H.}~\bibnamefont{Dosch}},
   \bibinfo{author}{\bibfnamefont{W.}~\bibnamefont{Drube}}, \bibnamefont{and}
   \bibinfo{author}{\bibfnamefont{B.~C.}~\bibnamefont{Cowie}},
   \bibinfo{journal}{Science} \textbf{\bibinfo{volume}{303}}
   (\bibinfo{year}{2004}) \bibinfo{pages}{1652} .
   
 \bibitem{park_etrij05}
   \bibinfo{author}{\bibfnamefont{S.-H.~K.}~\bibnamefont{Park}},
   \bibinfo{author}{\bibfnamefont{J.}~\bibnamefont{Oh}},
   \bibinfo{author}{\bibfnamefont{C.-S.}~\bibnamefont{Hwang}},
   \bibinfo{author}{\bibfnamefont{J.-I.}~\bibnamefont{Lee}},
   \bibinfo{author}{\bibfnamefont{Y.~S.}~\bibnamefont{Yang}},
   \bibinfo{author}{\bibfnamefont{H.~Y.}~\bibnamefont{Chu}}, \bibnamefont{and}
   \bibinfo{author}{\bibfnamefont{K.-Y.}~\bibnamefont{Kang}},
   \bibinfo{journal}{ETRI J.} \textbf{\bibinfo{volume}{27}}
   (\bibinfo{year}{2005}) \bibinfo{pages}{545} .
   
 \bibitem{chwang_apl03}
   \bibinfo{author}{\bibfnamefont{A.~B.}~\bibnamefont{Chwang}},
   \bibinfo{author}{\bibfnamefont{M.~A.}~\bibnamefont{Rothman}},
   \bibinfo{author}{\bibfnamefont{S.~Y.}~\bibnamefont{Mao}},
   \bibinfo{author}{\bibfnamefont{R.~H.}~\bibnamefont{Hewitt}},
   \bibinfo{author}{\bibfnamefont{M.~S.}~\bibnamefont{Weaver}},
   \bibinfo{author}{\bibfnamefont{J.~A.}~\bibnamefont{Silvernail}},
   \bibinfo{author}{\bibfnamefont{K.}~\bibnamefont{Rajan}},
   \bibinfo{author}{\bibfnamefont{M.}~\bibnamefont{Hack}},
   \bibinfo{author}{\bibfnamefont{J.~J.}~\bibnamefont{Brown}},
   \bibinfo{author}{\bibfnamefont{X.}~\bibnamefont{Chu}},
   \bibinfo{author}{\bibfnamefont{L.}~\bibnamefont{Moro}},
   \bibinfo{author}{\bibfnamefont{T.}~\bibnamefont{Krajewski}},
   \bibnamefont{and}
   \bibinfo{author}{\bibfnamefont{N.}~\bibnamefont{Rutherford}},
   \bibinfo{journal}{Appl. Phys. Lett.} \textbf{\bibinfo{volume}{83}}
   (\bibinfo{year}{2003}) \bibinfo{pages}{413} .
  
 \bibitem{voigt_mseb04}
   \bibinfo{author}{\bibfnamefont{M.}~\bibnamefont{Voigt}} \bibnamefont{and}
   \bibinfo{author}{\bibfnamefont{M.}~\bibnamefont{Sokolowski}},
   \bibinfo{journal}{Mater. Sci. Eng. B} \textbf{\bibinfo{volume}{109}}
   (\bibinfo{year}{2004}) \bibinfo{pages}{99} .
   
 \bibitem{krug_advp97} \bibinfo{author}{\bibfnamefont{J.}~\bibnamefont{Krug}},
   \bibinfo{journal}{Adv.  Phys.} \textbf{\bibinfo{volume}{46}}
   (\bibinfo{year}{1997}) \bibinfo{pages}{139}.

\bibitem{pimpinelli_bk}
\bibinfo{author}{\bibfnamefont{A.}~\bibnamefont{Pimpinelli}} \bibnamefont{and}
  \bibinfo{author}{\bibfnamefont{J.}~\bibnamefont{Villain}},
  \emph{\bibinfo{title}{Physics of {C}rystal {G}rowth}}, Monographs and Texts
  in Statistical Physics (\bibinfo{publisher}{Cambridge University Press},
  \bibinfo{year}{1999}).

\bibitem{barabasi-stanley_bk}
\bibinfo{author}{\bibfnamefont{A.-L.} \bibnamefont{Barab\'{a}si}}
  \bibnamefont{and} \bibinfo{author}{\bibfnamefont{H.~E.}
  \bibnamefont{Stanley}}, \emph{\bibinfo{title}{Fractal {C}oncepts in {S}urface
  {G}rowth}} (\bibinfo{publisher}{Cambridge University Press},
  \bibinfo{year}{1995}).
  
\bibitem{family_jpa85}
  \bibinfo{author}{\bibfnamefont{F.}~\bibnamefont{Family}} \bibnamefont{and}
  \bibinfo{author}{\bibfnamefont{T.}~\bibnamefont{Vicsek}},
  \bibinfo{journal}{J. Phys. A: Math. Gen.} \textbf{\bibinfo{volume}{18}}
  (\bibinfo{year}{1985}) \bibinfo{pages}{L75} .
  
\bibitem{family_pa90}
  \bibinfo{author}{\bibfnamefont{F.}~\bibnamefont{Family}},
  \bibinfo{journal}{Physica A} \textbf{\bibinfo{volume}{168}}
  (\bibinfo{year}{1990}) \bibinfo{pages}{561} .
  
\bibitem{family_jpa86}
  \bibinfo{author}{\bibfnamefont{F.}~\bibnamefont{Family}},
  \bibinfo{journal}{J.  Phys. A: Math. Gen.} \textbf{\bibinfo{volume}{19}}
  (\bibinfo{year}{1986}) \bibinfo{pages}{L441} .
  
\bibitem{kardar_prl86}
  \bibinfo{author}{\bibfnamefont{M.}~\bibnamefont{Kardar}},
  \bibinfo{author}{\bibfnamefont{G.}~\bibnamefont{Parisi}}, \bibnamefont{and}
  \bibinfo{author}{\bibfnamefont{Y.-C.} \bibnamefont{Zhang}},
  \bibinfo{journal}{Phys. Rev. Lett.} \textbf{\bibinfo{volume}{56}}
  (\bibinfo{year}{1986}) \bibinfo{pages}{889} .
  
\bibitem{wolf_epl90} \bibinfo{author}{\bibfnamefont{E.~D.} \bibnamefont{Wolf}}
  \bibnamefont{and} \bibinfo{author}{\bibfnamefont{J.}~\bibnamefont{Villain}},
  \bibinfo{journal}{Europhys. Lett.} \textbf{\bibinfo{volume}{13}}
  (\bibinfo{year}{1990}) \bibinfo{pages}{389} .
  
\bibitem{jullien_jpa85}
  \bibinfo{author}{\bibfnamefont{R.}~\bibnamefont{Jullien}} \bibnamefont{and}
  \bibinfo{author}{\bibfnamefont{R.}~\bibnamefont{Botet}},
  \bibinfo{journal}{J.  Phys. A: Math. Gen.} \textbf{\bibinfo{volume}{18}}
  (\bibinfo{year}{1985}) \bibinfo{pages}{2279} .
  
\bibitem{krug_prl91} \bibinfo{author}{\bibfnamefont{J.}~\bibnamefont{Krug}}
  \bibnamefont{and} \bibinfo{author}{\bibfnamefont{P.}~\bibnamefont{Meakin}},
  \bibinfo{journal}{Phys. Rev. Lett.} \textbf{\bibinfo{volume}{66}}
  (\bibinfo{year}{1991}) \bibinfo{pages}{703} .
  
\bibitem{lai_prl91} \bibinfo{author}{\bibfnamefont{Z.-W.} \bibnamefont{Lai}}
  \bibnamefont{and} \bibinfo{author}{\bibfnamefont{S.~D.}
    \bibnamefont{Sarma}}, \bibinfo{journal}{Phys. Rev. Lett.}
  \textbf{\bibinfo{volume}{66}} (\bibinfo{year}{1991}) \bibinfo{pages}{2348} .
  
\bibitem{tang_prl91} \bibinfo{author}{\bibfnamefont{L.-H.} \bibnamefont{Tang}}
  \bibnamefont{and}
  \bibinfo{author}{\bibfnamefont{T.}~\bibnamefont{Nattermann}},
  \bibinfo{journal}{Phys. Rev. Lett.} \textbf{\bibinfo{volume}{66}}
  (\bibinfo{year}{1991}) \bibinfo{pages}{2899} .
  
\bibitem{duerr_prl03} \bibinfo{author}{\bibfnamefont{A.~C.}
    \bibnamefont{D{\"u}rr}},
  \bibinfo{author}{\bibfnamefont{F.}~\bibnamefont{Schreiber}},
  \bibinfo{author}{\bibfnamefont{K.~A.} \bibnamefont{Ritley}},
  \bibinfo{author}{\bibfnamefont{V.}~\bibnamefont{Kruppa}},
  \bibinfo{author}{\bibfnamefont{J.}~\bibnamefont{Krug}},
  \bibinfo{author}{\bibfnamefont{H.}~\bibnamefont{Dosch}}, \bibnamefont{and}
  \bibinfo{author}{\bibfnamefont{B.}~\bibnamefont{Struth}},
  \bibinfo{journal}{Phys. Rev. Lett.} \textbf{\bibinfo{volume}{90}}
  (\bibinfo{year}{2003}) \bibinfo{pages}{016104}.
  
\bibitem{he_prl92} \bibinfo{author}{\bibfnamefont{Y.-L.} \bibnamefont{He}},
  \bibinfo{author}{\bibfnamefont{H.-N.} \bibnamefont{Yang}},
  \bibinfo{author}{\bibfnamefont{T.-M.} \bibnamefont{Lu}}, \bibnamefont{and}
  \bibinfo{author}{\bibfnamefont{G.-C.} \bibnamefont{Wang}},
  \bibinfo{journal}{Phys. Rev. Lett.} \textbf{\bibinfo{volume}{69}}
  (\bibinfo{year}{1992}) \bibinfo{pages}{3770} .
  
\bibitem{thompson_prb94}
  \bibinfo{author}{\bibfnamefont{C.}~\bibnamefont{Thompson}},
  \bibinfo{author}{\bibfnamefont{G.}~\bibnamefont{Palasantzas}},
  \bibinfo{author}{\bibfnamefont{Y.~P.} \bibnamefont{Feng}},
  \bibinfo{author}{\bibfnamefont{S.~K.} \bibnamefont{Sinha}},
  \bibnamefont{and} \bibinfo{author}{\bibfnamefont{J.}~\bibnamefont{Krim}},
  \bibinfo{journal}{Phys. Rev. B} \textbf{\bibinfo{volume}{49}}
  (\bibinfo{year}{1994}) \bibinfo{pages}{4902}.
  
\bibitem{you_prl93} \bibinfo{author}{\bibfnamefont{H.}~\bibnamefont{You}},
  \bibinfo{author}{\bibfnamefont{R.~P.} \bibnamefont{Chiarello}},
  \bibinfo{author}{\bibfnamefont{H.-K.} \bibnamefont{Kim}}, \bibnamefont{and}
  \bibinfo{author}{\bibfnamefont{K.~G.} \bibnamefont{Vandervoort}},
  \bibinfo{journal}{Phys. Rev. Lett.} \textbf{\bibinfo{volume}{70}}
  (\bibinfo{year}{1993}) \bibinfo{pages}{2900} .
  
\bibitem{jun_epl98} \bibinfo{author}{\bibfnamefont{J.}~\bibnamefont{Wang}},
  \bibinfo{author}{\bibfnamefont{G.}~\bibnamefont{Li}},
  \bibinfo{author}{\bibfnamefont{P.}~\bibnamefont{Yang}},
  \bibinfo{author}{\bibfnamefont{M.}~\bibnamefont{Cui}},
  \bibinfo{author}{\bibfnamefont{X.}~\bibnamefont{Jiang}},
  \bibinfo{author}{\bibfnamefont{B.}~\bibnamefont{Dong}}, \bibnamefont{and}
  \bibinfo{author}{\bibfnamefont{H.}~\bibnamefont{Liu}},
  \bibinfo{journal}{Europhys. Lett.} \textbf{\bibinfo{volume}{42}}
  (\bibinfo{year}{1998}) \bibinfo{pages}{283} .

\bibitem{auger_jap05} \bibinfo{author}{\bibfnamefont{M.~A.}~\bibnamefont{Auger}},
  \bibinfo{author}{\bibfnamefont{L.}~\bibnamefont{V\'azquez}},
  \bibinfo{author}{\bibfnamefont{O.}~\bibnamefont{S\'anchez}},
  \bibinfo{author}{\bibfnamefont{M.}~\bibnamefont{Jergel}},
  \bibinfo{author}{\bibfnamefont{R.}~\bibnamefont{Cuerno}}, \bibnamefont{and}
  \bibinfo{author}{\bibfnamefont{M.}~\bibnamefont{Castro}},
  \bibinfo{journal}{J. Appl. Phys.} \textbf{\bibinfo{volume}{97}}
  (\bibinfo{year}{2005}) \bibinfo{pages}{123528} .
  
\bibitem{lindstroem_tsf01}
  \bibinfo{author}{\bibfnamefont{T.}~\bibnamefont{Lindstr\"om}},
  \bibinfo{author}{\bibfnamefont{J.}~\bibnamefont{Isidorsson}},
  \bibnamefont{and}
  \bibinfo{author}{\bibfnamefont{G.~A.}~\bibnamefont{Niklasson}},
  \bibinfo{journal}{Thin Solid Films} \textbf{\bibinfo{volume}{401}}
  (\bibinfo{year}{2001}) \bibinfo{pages}{165} .
  
\bibitem{collins_prl94} \bibinfo{author}{\bibfnamefont{G.~W.}
    \bibnamefont{Collins}}, \bibinfo{author}{\bibfnamefont{S.~A.}
    \bibnamefont{Letts}}, \bibinfo{author}{\bibfnamefont{E.~M.}
    \bibnamefont{Fearon}}, \bibinfo{author}{\bibfnamefont{R.~L.}
    \bibnamefont{McEachern}}, \bibnamefont{and}
  \bibinfo{author}{\bibfnamefont{T.~P.}  \bibnamefont{Bernat}},
  \bibinfo{journal}{Phys. Rev. Lett.}  \textbf{\bibinfo{volume}{73}}
  (\bibinfo{year}{1994}) \bibinfo{pages}{708}.
  
\bibitem{ernst_prl94} \bibinfo{author}{\bibfnamefont{H.-J.}
    \bibnamefont{Ernst}},
  \bibinfo{author}{\bibfnamefont{F.}~\bibnamefont{Fabre}},
  \bibinfo{author}{\bibfnamefont{R.}~\bibnamefont{Folkerts}},
  \bibnamefont{and}
  \bibinfo{author}{\bibfnamefont{J.}~\bibnamefont{Lapujoulade}},
  \bibinfo{journal}{Phys. Rev. Lett.} \textbf{\bibinfo{volume}{72}}
  (\bibinfo{year}{1994}) \bibinfo{pages}{112}.
  
\bibitem{duerr_apl02} \bibinfo{author}{\bibfnamefont{A.~C.}
    \bibnamefont{D{\"u}rr}},
  \bibinfo{author}{\bibfnamefont{F.}~\bibnamefont{Schreiber}},
  \bibinfo{author}{\bibfnamefont{M.}~\bibnamefont{M{\"u}nch}},
  \bibinfo{author}{\bibfnamefont{N.}~\bibnamefont{Karl}},
  \bibinfo{author}{\bibfnamefont{B.}~\bibnamefont{Krause}},
  \bibinfo{author}{\bibfnamefont{V.}~\bibnamefont{Kruppa}}, \bibnamefont{and}
  \bibinfo{author}{\bibfnamefont{H.}~\bibnamefont{Dosch}},
  \bibinfo{journal}{Appl. Phys. Lett.} \textbf{\bibinfo{volume}{81}}
  (\bibinfo{year}{2002}) \bibinfo{pages}{2276}.
  
\bibitem{kowarik_prl06} \bibinfo{author}{\bibfnamefont{S.}
    \bibnamefont{Kowarik}},
  \bibinfo{author}{\bibfnamefont{A.}~\bibnamefont{Gerlach}},
  \bibinfo{author}{\bibfnamefont{S.}~\bibnamefont{Sellner}},
  \bibinfo{author}{\bibfnamefont{F.}~\bibnamefont{Schreiber}},
  \bibinfo{author}{\bibfnamefont{L.}~\bibnamefont{Cavalcanti}},
  \bibnamefont{and}
  \bibinfo{author}{\bibfnamefont{O.}~\bibnamefont{Konovalov}},
  \bibinfo{journal}{Phys. Rev. Lett.} \textbf{\bibinfo{volume}{96}}
  (\bibinfo{year}{2006}) \bibinfo{pages}{125504}.
  
\bibitem{duerr_am02} \bibinfo{author}{\bibfnamefont{A.~C.}
    \bibnamefont{D{\"u}rr}},
  \bibinfo{author}{\bibfnamefont{F.}~\bibnamefont{Schreiber}},
  \bibinfo{author}{\bibfnamefont{M.}~\bibnamefont{M{\"u}nch}},
  \bibinfo{author}{\bibfnamefont{B.} \bibnamefont{Krause}},
  \bibinfo{author}{\bibfnamefont{V.} \bibnamefont{Kruppa}}, \bibnamefont{and}
  \bibinfo{author}{\bibfnamefont{H.}~\bibnamefont{Dosch}},
  \bibinfo{journal}{Adv. Mater.} \textbf{\bibinfo{volume}{14}}
  (\bibinfo{year}{2002}) \bibinfo{pages}{961}.
  
\bibitem{sellner_am04}
  \bibinfo{author}{\bibfnamefont{S.}~\bibnamefont{Sellner}},
  \bibinfo{author}{\bibfnamefont{A.}~\bibnamefont{Gerlach}},
  \bibinfo{author}{\bibfnamefont{F.}~\bibnamefont{Schreiber}},
  \bibinfo{author}{\bibfnamefont{M.}~\bibnamefont{Kelsch}},
  \bibinfo{author}{\bibfnamefont{N.}~\bibnamefont{Kasper}},
  \bibinfo{author}{\bibfnamefont{H.}~\bibnamefont{Dosch}},
  \bibinfo{author}{\bibfnamefont{S.}~\bibnamefont{Meyer}},
  \bibinfo{author}{\bibfnamefont{J.}~\bibnamefont{Pflaum}},
  \bibinfo{author}{\bibfnamefont{M.}~\bibnamefont{Fischer}}, \bibnamefont{and}
  \bibinfo{author}{\bibfnamefont{B.}~\bibnamefont{Gompf}},
  \bibinfo{journal}{Adv. Mater.} \textbf{\bibinfo{volume}{16}}
  (\bibinfo{year}{2004}) \bibinfo{pages}{1750}.
  
\bibitem{sellner_LPaper}
  \bibinfo{author}{\bibfnamefont{S.}~\bibnamefont{Sellner}},
  \bibinfo{author}{\bibfnamefont{A.}~\bibnamefont{Gerlach}},
  \bibinfo{author}{\bibfnamefont{F.}~\bibnamefont{Schreiber}},
  \bibinfo{author}{\bibfnamefont{M.}~\bibnamefont{Kelsch}},
  \bibinfo{author}{\bibfnamefont{N.}~\bibnamefont{Kasper}},
  \bibinfo{author}{\bibfnamefont{H.}~\bibnamefont{Dosch}},
  \bibinfo{author}{\bibfnamefont{S.}~\bibnamefont{Meyer}},
  \bibinfo{author}{\bibfnamefont{J.}~\bibnamefont{Pflaum}},
  \bibinfo{author}{\bibfnamefont{M.}~\bibnamefont{Fischer}},
  \bibinfo{author}{\bibfnamefont{B.}~\bibnamefont{Gompf}}, \bibnamefont{and}
  \bibinfo{author}{\bibfnamefont{G.}~\bibnamefont{Ulbricht}},
  \bibinfo{journal}{J. Mater. Res.} \textbf{\bibinfo{volume}{21}}
  (\bibinfo{year}{2006}) \bibinfo{pages}{455}.
  
\bibitem{meyer_submitted}
  \bibinfo{author}{\bibfnamefont{S.}~\bibnamefont{Meyer}},
  \bibinfo{author}{\bibfnamefont{S.}~\bibnamefont{Sellner}},
  \bibinfo{author}{\bibfnamefont{F.}~\bibnamefont{Schreiber}},
  \bibinfo{author}{\bibfnamefont{H.}~\bibnamefont{Dosch}},
  \bibinfo{author}{\bibfnamefont{G.}~\bibnamefont{Ulbricht}},
  \bibinfo{author}{\bibfnamefont{M.}~\bibnamefont{Fischer}},
  \bibinfo{author}{\bibfnamefont{B.}~\bibnamefont{Gompf}}, \bibnamefont{and}
  \bibinfo{author}{\bibfnamefont{J.}~\bibnamefont{Pflaum}},
  \bibinfo{journal}{Mater. Res. Soc. Symp. Proc.}
  \textbf{\bibinfo{volume}{965E}} (\bibinfo{year}{2007}) \bibinfo{pages}{6}.
  
\bibitem{parratt_pr54} \bibinfo{author}{\bibfnamefont{L.~G.}
    \bibnamefont{Parratt}}, \bibinfo{journal}{Phys. Rev.}
  \textbf{\bibinfo{volume}{95}} (\bibinfo{year}{1954}) \bibinfo{pages}{359}.
  
\bibitem{schlomka_prb95} \bibinfo{author}{\bibfnamefont{J.-P.}
    \bibnamefont{Schlomka}},
  \bibinfo{author}{\bibfnamefont{M.}~\bibnamefont{Tolan}},
  \bibinfo{author}{\bibfnamefont{L.}~\bibnamefont{Schwalowsky}},
  \bibinfo{author}{\bibfnamefont{O.~H.} \bibnamefont{Seeck}},
  \bibinfo{author}{\bibfnamefont{J.}~\bibnamefont{Stettner}},
  \bibnamefont{and} \bibinfo{author}{\bibfnamefont{W.}~\bibnamefont{Press}},
  \bibinfo{journal}{Phys. Rev. B} \textbf{\bibinfo{volume}{51}}
  (\bibinfo{year}{1995}) \bibinfo{pages}{2311}.

\bibitem{holy_bk}
\bibinfo{author}{\bibfnamefont{V.}~\bibnamefont{Hol{\'y}}},
  \bibinfo{author}{\bibfnamefont{U.}~\bibnamefont{Pietsch}}, \bibnamefont{and}
  \bibinfo{author}{\bibfnamefont{T.}~\bibnamefont{Baumbach}},
  \emph{\bibinfo{title}{High-Resolution {X}-Ray {S}cattering from {T}hin
  {F}ilms and {M}ultilayers}} (\bibinfo{publisher}{Springer Berlin},
  \bibinfo{address}{Heidelberg, New York}, \bibinfo{year}{1999}).

\end{thebibliography}
\end{document}